\documentclass[12pt]{article}
\usepackage{amsgen,amsmath,amstext,amsbsy,amsopn,amsfonts,amssymb}
\usepackage{amsgen,amstext,amsbsy,amsopn,amsfonts,amssymb}
\usepackage[pdftex]{graphicx}
\usepackage{hyperref}
\usepackage{graphpap}
\usepackage{rotating}
\usepackage{url}

\def\PF{PowerFlux~}

\def\sci#1#2{{#1}{\times 10^{#2}}}

\def\expect{\mathbb E}
\def\Torus{{\mathbb T}}

\def\SNR{{\textrm{SNR}}}

\def\address#1{ {#1} }
\def\ead#1{E-mail: \url{#1} }
\def\pacs#1{PACS numbers: {#1} }
\def\lineup{\vskip 0.5cm}
\def\br{\hline}
\def\mr{\hline}
\def\0{~}

\title{On blind searches for noise dominated signals: a loosely coherent approach.}
\author{Vladimir Dergachev}

\begin{document}

\maketitle
\address{
LIGO Laboratory,
California Institute of Technology,
MS 18-34,
Pasadena, CA 91125, USA
}
\ead{volodya@caltech.edu}
\begin{abstract}
 We introduce a ``loosely coherent'' method for detection of continuous gravitational waves that bridges the gap between semi-coherent and purely coherent methods. Explicit control over accepted families of signals is used to increase sensitivity of power-based statistic while avoiding the high computational costs of conventional matched filters. Several examples as well as a prototype implementation are discussed.
\end{abstract}


\section{Introduction}

The need for methods described in this paper arose during development of the \PF search \cite{S4IncoherentPaper} for continuous gravitational wave signals. Even though aimed at a specific purpose of following up \PF outliers, they have much wider applicability. To that end we will present a simplified description that omits some technicalities specific to searches for continuous gravitational waves.

The \PF algorithm \cite{PowerFluxTechNote} detects gravitational waves by computing power received from a particular direction at a certain frequency and spindown. Similar approaches include Hough and StackSlide  searches \cite{S4IncoherentPaper,BCCS,BC00,cgk,StackSlideTechNote,S2HoughPaper,hough04}. Also, searches have been carried out with algorithms using substantially larger coherence lengths such as $\mathcal{F}$-statistic \cite{S2FstatPaper,S4EH,S5EH}.

The power-based methods are computationally efficient and allow all-sky blind searches to be performed with the sensitivity scaling as fourth root of the amount $N$ of analyzed data. In contrast, coherent searches scale as $N^{-1/2}$ but become impractical 
for moderate values of $N$. They also rely heavily on knowing the exact form of the expected signal - an assumption that we feel is overly bold when one is looking for a form of radiation for which no prior direct observation exists.

There are searches that fill the space between these extremes. One way is to combine incoherently an output of multiple coherent searches. Another approach is to perform a hierarchical search that follows up outliers with longer baseline coherent investigation. Both employ longer coherence baselines than power-based methods.

Thus, in order to make a successful detection, one needs to overcome a ``potential barrier'' in computational costs that separates a blind search from an easy verification of a successful candidate.

One reason for difficulties with current coherent methods is that they are optimized with a specific signal waveform in mind, and then the search is iterated over many signal templates. The templates often overlap \cite{Fmetric} and, in fact, oversampling is routinely used to ensure that no signals are missed. This design is well warranted if sufficient computational power exists to exhaust the entire search space - but this is a situation current gravitational wave searches are {\bf not} in. Furthermore, maximization alone is not necessarily the most optimal statistic \cite{LargeCorr, BayesianFstat}.

We believe that an approach that combines attention to sensitivity and computational efficiency with more agile control over accepted waveforms is both more physically prudent and computationally accessible. To illustrate this, we present a {\em loosely coherent} method that is based on estimating power for a family of signal waveforms at once.

\section{Statement of the problem}

For the purposes of this paper we will assume that our entire dataset has been broken up into $N$ short portions each of which has been subjected to the Fourier transform, and we are looking for a signal of constant amplitude that would land into a single frequency bin $\left\{a_k\right\}_{k=1}^{N}$ in $k$-th short Fourier transform with varying phases $\left\{\phi_k\right\}_{k=1}^{N}$. 

If the phases were known in advance we could compute the power of a coherent sum
\begin{equation}
\label{eqn:p_equation}
P=\left|\sum_{k=1}^{N} a_k e^{-i \phi_k}\right|^2=\sum_{k,l=1}^{N} a_k^*a_l e^{i (\phi_k-\phi_l)}
\end{equation}
the high values of which would indicate the presence of the signal. There is a large body of literature that describes designing statistics with optimal signal-to-noise ratios (SNR), in particular \cite{WZ}. 

In many cases a part of signal evolution (such as Doppler modulation induced by motion of the Earth) is known in advance. If we assume that this contribution has been factored out then the coherent power sum reduces to the case $\phi_{l+1} - \phi_l=0$.

The set $\Torus$ of all possible phases (modulo $2\pi$) forms an $N$-dimensional torus on which $P$ is a smooth function. In practice, phases cannot be determined exactly ahead of time, but rather obey a set of constraints. Such family of signals would sweep a submanifold $S \subset \Torus$, possibly with boundary.

Our goal is then to find a statistic that achieves high values when a signal from $S$ is present and low values otherwise. One way to do that is take the maximum of $P$ over $\left\{\phi_k\right\}$ constrained to the submanifold $S$. Another approach is to view the unknown parameters as random, with the phases forming stochastic process, usually highly correlated. It is important to note that for either detection or establishment of upper limits we only need to know whether the signal is present, as the parameter estimation can be performed by partitioning $S$ into subsets.

We call this a {\em loosely coherent} approach, as instead of trying to find signals with a certain pre-determined set of phases, we are content with any signal that has phase evolution from $S$. The choice of the set $S$ and the statistic $P$ is then up to the designer of the search thus providing the necessary freedom to satisfy conflicting demands of efficiency in computation and signal recovery.

Of course, any practical detection algorithm, even designed with full knowledge of expected signal, will respond to data with signals from a wider set of phases than physically expected. Tailoring the set $S$ at the design stage, rather than simply characterizing it after implementation, allows finer control over which astrophysical signals one can detect and particulars of template placement.

\section{Implementation of loosely coherent statistics}

\subsection{Maximization}
The most straightforward way to construct a loosely coherent statistic is to maximize $P$ over the set of possible phases $S$.
This is a classical optimization problem with a quadratic objective that possesses several difficulties:

First, we are trying to {\em maximize} a non-negative definite quadratic function - thus our problem is inherently non-convex\footnote{A maximization problem $\max_{x\in S} f(x)$ is called convex if the set of points $\left\{(x,y): x\in S\textrm{ and } y\le f(x)\right\}$ is convex. In particular, for a differentiable $f$, this assures that the gradient descent method can not become stuck in a valley.}, even for small portions of $S$. This precludes the use of well known optimization methods like gradient descent.

Secondly, the dimension $N$ is very large, with small searches starting at $N=1000$.

The third difficulty is more subtle and is due to the nature of interesting signal families $S$. These usually involve phases that evolve moderately fast with $k$ and can wrap around numerous times. A typical example is a linear evolution produced by mismatch in frequency given by 
\begin{equation}
\phi_k=A+Bk
\end{equation}
with $B$ on the order of $0.1$.

Because of the wrap around, a small uncertainty in $\phi_k$ for some $k$ can result in very large uncertainty in $\phi_l$ for $|l-k|\gg 1$. In the limit $N\rightarrow \infty$ the embedding of $S$ into the torus $\Torus^\infty$ (considered with $L^\infty$ norm in which it is not compact) stops being differentiable or continuous altogether.

The properties of the map $\phi: S\rightarrow \Torus$ as $N$ approaches infinity are tightly connected with the scalability in the number of templates. To describe this connection we need some well-known tools from functional analysis.

Let $S$ be a bounded (i.e. compact) finite dimensional manifold, possibly with boundary, with a metric $\rho_S$. As mentioned before, we consider the torii $\Torus^N$ with $L^\infty$ metric 
\begin{equation}
\rho_N(\{\phi_k\}_{k=1}^N, \{\psi_k\}_{k=1}^N)=\sup_k \inf_m \left|\phi_k-\psi_k-2\pi m\right|
\end{equation}
Let $\Phi_N: S\rightarrow \Torus^N$ be the family of embeddings describing phase evolution for successive SFTs.

Our goal is to select templates in $S$ such their image under $\Phi_N$ forms an $\epsilon$-net - any point in $\Phi_N(S)$ is within $\epsilon$ of an image of some template.

We distinguish three fundamentally different situations:
\begin{itemize}
 \item The map $\Phi_\infty: S\rightarrow \Torus^\infty$ is Lipschitz, i.e. it satisfies the following property:
\begin{equation}
\rho_\infty(\Phi(x_0), \Phi_\infty(x_1))< L \rho_S(x_0, x_1)
\end{equation}
Any continuously differentiable map is Lipschitz.
In this case, we can cover $\Phi_\infty(S)$ with any desired tolerance $\epsilon$ by constructing a set of templates in $S$ which forms an $\epsilon/L$-net. A well-known fact from topology \cite{dimension_theory} is that it is possible to find coverings with template count scaling as $\epsilon^{-d}$ where $d$ is the Hausdorff dimension of $S$.

Thus, we see that the template count does not depend on $N$ and is proportional to $\epsilon^{-\dim(S)}$ - the best we could hope for. An example of such a map is given by
\begin{equation}
\Phi_\infty(A, B)=\{A \sin(\omega k+B)\}_{k=1}^{\infty}
\end{equation}
where $\omega$ is a fixed parameter (such as Earth rotation frequency) and $A$ and $B$ are bounded search parameters. A physically relevant example is given by phase shifts from amplitude response of the detector.

\item The map $\Phi_\infty: S\rightarrow \Torus^\infty$ is known to be continuous, but not Lipschitz. In this case, we can still find a suitable template set for any desired tolerance $\epsilon$, but the spacing of the templates in $S$ will not depend linearly on $\epsilon$ as it does in the Lipschitz case. We thus retain independence of $N$ but the number of required templates can grow faster than $\epsilon^{-\dim(S)}$. 

A mathematical example of such a map is given by
\begin{equation}
\Phi(A)=\left\{\frac{\sin(A k)}{\sqrt{k}}\right\}_{k=1}^{\infty}
\end{equation}
The required template count grows as $\epsilon^{-2}$. We are not aware of any physically motivated search for continuous gravitational radiation that has parameters of this form.

\item The map $\Phi_\infty: S\rightarrow \Torus^\infty$ is not continuous. While this can be due to trivial causes such as partial breaks in otherwise Lipschitz map, in general it would not be possible to find a finite template set to cover $\Torus^\infty$. For the finite case $\Phi_\infty: S\rightarrow \Torus^N$ the template count will grow with $N$.

An example of such a map is given by frequency evolution discussed above:
\begin{equation}
\Phi_N(A)=\left\{Ak\right\}_{k=1}^{N}
\end{equation}
for which the required template count scales as $N\epsilon^{-1}$. 
\end{itemize}

One way to deal with these difficulties is to partition $N$ into small enough sets so that maximization can actually be carried out and combine the results afterwards. Further computational savings result from picking $S$ described by only a few necessary parameters and overcoming their scaling properties with large computing power.
The coherent searches for gravitational radiation such as \cite{S2FstatPaper,S4EH,S5EH} can be viewed as examples
of this approach.

\subsection{Averaging}
Another way to bring computational costs under control is to replace $P$ with a related function with a smaller Lipschitz
constant. One can achieve this by averaging $P$ over $S$ or its subsets, which is equivalent to computing expectation value of $P$ over some assumed distribution on $S$. This spreads the signal response over a larger area, but we only have to make the computation once for each subset. For ease of exposition we use the usual Lebesgue measure and average power rather than a more complicated statistic such as likelihood.

In the most extreme case we just average away the phases $\phi_k$ yielding the conventional semi-coherent method:
\begin{equation}
\expect P=\expect \sum_{k,l=1}^{N} a_k^*a_l e^{i (\phi_k-\phi_l)} = \sum_{l=1}^{N} |a_l|^2
\end{equation}
If the phases are truly random this statistic will perform better in the presence of well behaved noise than computation of the maximum \cite{WZ}.

A more conservative approach will limit phase evolution:
\begin{equation}
S=\left\{\left\{\phi_k\right\} : |\phi_l-\phi_{l+1}|<\delta\right\}
\end{equation}
yielding the following statistic (computed using variables $\delta_l=\phi_{l+1}-\phi_l$):
\begin{equation}
\expect P = \frac{1}{(2\delta)^{N-1}}\int_{-\delta}^{\delta} \dots \int_{-\delta}^{\delta} P d\delta_1 \dots d\delta_{N-1}=\sum_{k,l=1}^{N} a_k^*a_l \left(\frac{\sin(\delta)}{\delta}\right)^{|k-l|}
\end{equation}
which interpolates between the fully coherent sum for $\delta=0$ and the semi-coherent case $\delta=\pi$. The allowed spacing between frequency templates increases with $\delta$, and in the limiting case $N\rightarrow \infty$ is determined by the value of $\delta/\pi$ in units of frequency bins. This has proven to be a good initial estimate of the spacing required by searches where $N\gg 1/\delta$.

This method will lose some power if the true frequency of the signal at the time corresponding to coefficient $a_k$ is not a harmonic sampled by the Fourier transform. To avoid this, one can replace $a_k$ with more precise values estimated from the Dirichlet kernel. This effectively makes sure that the point with all phases $0$ belongs to $S$ - a condition we assume from now on.

It is also possible to use the same approach to reduce the influence of periodic changes of underlying frequency, such as caused by mismatch in sky position and the resultant Doppler shifts. Assume
\begin{equation}
S=\left\{\left\{\phi_k\right\} : \phi_k=A+B\sin(\omega k+C)\right\}
\end{equation}
where $A$ is some unknown (and irrelevant) phase, $\omega$ and $C$ are known and fixed (such as from sidereal Doppler modulation) and $B$ is allowed to vary, subject to $|B|\leq \beta$.
Then
\begin{equation}
\begin{array}{l}
\expect P=\frac{1}{2\beta}\int_{-\beta}^{\beta} \sum_{k,l=1}^{N} a_k^*a_l e^{i (\phi_k-\phi_l)} dB = \\
\quad\quad\quad=\sum_{k,l=1}^{N} a_k^*a_l \frac{\sin(\beta( \sin(\omega k+C)-\sin(\omega l+C)))}{\beta( \sin(\omega k+C)-\sin(\omega l+C))}=\\
\quad\quad\quad=\sum_{k,l=1}^{N} a_k^*a_l \frac{\sin(2\beta\sin(\omega (k-l)/2)\cos(C))}{2\beta \sin(\omega (k-l)/2)\cos(C)}
\end{array}
\end{equation}

As we have chosen a simple power sum $P$ as a starting point, our averaged statistic will always have the form
\begin{equation}
\label{eqn:kernel_statistic}
\expect P = \sum_{k,l=1}^{N} a_k^*a_l K_{kl}
\end{equation}
for some kernel $K_{kl}$ and is thus similar to cross-correlation search \cite{cross_correllation}. As we will see later the efficient computation of the sum for small $\delta$ is best done in a manner different from the cross-correlation statistic.

\subsection{Loosely coherent searches as a filtering problem}

The statistic $\expect P$ can be rewritten as a scalar product of the vector of input data $a$ with the image of $a$ under the operator $K'$ which square $\bar{K'}^tK'$ is given by the kernel $K_{kl}$:
\begin{equation}
\expect P = \sum_{k,l=1}^{N} a_k^*a_l K_{kl}= \bar{a}^t\bar{K'}^t K'a
\end{equation}
From this point of view $K'$ acts as a filter rejecting signals outside the expected set, after which we take the usual semi-coherent sum.

For example, $K'$ can be chosen as a low-pass filter given by a $sinc$ or Lanczos kernel. This would admit signals with phases varying slower than the filter cutoff frequency. 

For a practical implementation the main point of concern is the ability of the statistic to tolerate frequency mismatch, as it directly impacts the number of templates. For this purpose the low pass filters are optimum, tolerating mismatch values up to a cutoff frequency and rejecting signals with faster varying phases.

A more sophisticated approach is to assume a distribution on the set of allowed phases and then treat our signal as a highly correlated stochastic process. Since the data analysis is typically carried out after the data collection is complete, one is not restricted to causal filters alone and, in the case of stationary noise and limited phase evolution, we obtain a low pass filter as a solution.

The loosely coherent statistic based on a $sinc$ filter is optimal in the following idealized situation: suppose our data $\left\{a_k\right\}$ consists of a sum of stationary mean zero Gaussian noise of known variation (which is typically easy to estimate from data known not to contain any signals) and unknown band-limited signal of limited power, with no additional information on the signal form or phase evolution. A Fourier transform will separate our data into high frequency area where there is no signal and which can be safely discarded and low frequency area which phase information is irrelevant due to the signal having an arbitrary spectral shape.

We are thus left with a problem of deciding whether our low-frequency data is consistent with Gaussian noise alone or there is an arbitrary additive signal present. 

Both the limited power condition and the structure of Gaussian noise are symmetric under unitary transformations. Thus, if no other restrictions are present, the only meaningful information is the power contained in the low-frequency data.

While this fairly standard argument bridges both frequentist and Bayesian approaches, it does have a number of limitations. The most severe is that the symmetry is lost in case of non-stationary noise. Additionally, a family of physical signals can be expected to have a spectrum more interesting than a plain flat-top.

\section{Practicalities of the gravitational wave searches}
We will now qualify the phase shift evolution that one expects to encounter in current searches.

At the moment, the searches analyze data from $50$~Hz through $1500$~Hz, accounting for spindowns as large as $-10^{-8}$~Hz/s.
The analysis is done using short Fourier transforms (SFTs) of $\approx 1800$~s length, which have $50\%$ overlap in some searches, no overlap in others and often have gaps. For this paper we will assume that the time interval $\Delta t$ between $a_k$ and $a_{k+1}$ is $1800$~s.

We will assume that $\left\{a_k\right\}$ have already been adjusted so that the template $O$ with all $\phi_k=0$ is in $S$.

There are several sources of non-trivial phase shifts, which we will describe in terms of maximum expected difference $\delta$ between nearby phases:
\begin{itemize}
 \item Frequency mismatch - a template possessing frequency different from $O$ by $\Delta f$ will experience a linear phase evolution of 
\begin{equation}
\delta=2\pi \Delta f \Delta t
\end{equation}
\item Sky position mismatch - a mismatch in sky position will produce a slightly different Doppler shift. On short time scales this is dominated by Earth rotation (with velocity $\sim\sci{1}{-6}$c) and is periodic in time and linear in sampled frequency: 
\begin{equation}
\delta=2\pi 10^{-6} f \Delta t \Delta r
\end{equation}
where $\Delta r$ is the maximum expected mismatch in radians, with practical values usually less than $0.01$.
\item Spindown mismatch - a spindown different from $O$ by $\Delta \dot{f}$ will produce a linear evolution of the frequency and, thus, a quadratic change in phase:
\begin{equation}
\delta=2\pi \Delta \dot{f} T \Delta t
\end{equation}
Here $T$ shows maximum variation of time variable with respect to reference time. If the reference time is positioned at the center of the run, then $T$ is half the time base.

\item Source frequency evolution - the source signal can be modulated by a nearby orbiting object. Assuming circular orbit with radius $r$ (expressed in astronomical units) and using $\rho=m/M$ for the ratio of object mass $m$ to the star mass $M$ (both expressed in units of solar mass) the angular frequency of the modulation is:
\begin{equation}
\omega=\sqrt{\frac{G(m+M)}{r^3}}\approx 2\times 10^{-7} \textrm{~Hz} \cdot \sqrt{\frac{M(1+\rho)}{r^3}}
\end{equation}
and the maximum Doppler shift from the central body is
\begin{equation}
\frac{v}{c}\approx \frac{\rho}{c}\sqrt{\frac{G(m+M)}{r}}\approx 10^{-4} \rho \sqrt{\frac{M}{r}}
\end{equation}
The worst case change in phase induced by this motion over time $\Delta t$ and assuming radiating frequency $f$ is:
\begin{equation}
\delta=f \frac{2v}{c} \omega \Delta t \approx 2.2\times 10^{-5} \textrm{~Hz} \cdot \frac{f}{1000 \textrm{~Hz}} \frac{\Delta t}{1800 \textrm{~s}} \frac{M/M_{\textrm{SUN}}}{(r/{1\textrm{~AU}})^2} \rho \sqrt{1+\rho}
\end{equation}
The curiously small size of $\delta$ is due largely to the small value of product $\omega \Delta t$. For a search that assumes a specific phase evolution over a long time interval this would be much larger. The loosely coherent search is not completely immune from this effect - it will lose power when enough phase accumulates during integration for the signal to escape into nearby frequency bin. This suggests that searches looking for more extreme systems should  use coarser frequency bins, smaller $\Delta t$ and tighter $\delta$.
\end{itemize}

Table \ref{tab:phase_shift} shows the expected phase shift for conditions commonly encountered in present day searches.

\begin{table}[hbtp]
\begin{center}
\caption{Maximum phase change in degrees between $1800$~second spaced SFTs.}
\lineup
\label{tab:phase_shift}
\begin{tabular}{@{}p{6cm}lllll}
\br
Phase shift cause & $100$~Hz & $500$~Hz & $1000$~Hz & $2000$~Hz\\
\mr
Frequency mismatch of\par $\Delta f=0.1/\Delta t$ & $36$ & $36$ & $36$ & $36$\\
Sky position mismatch of\par $\Delta r=1^\circ$  & \0$1.1$ & \0$6$ & $11$ & $23$ \\
Spindown mismatch of\par $\Delta \dot{f}=10^{-12}$~Hz/s for $T=1$~y & $20$ & $20$  & $20$  & $20$ \\
Source modulation for\par $\rho=1$ and $r=0.1$~AU & \0$0.1$ & \0$0.6$ & \0$1$ & \0$2$\\
\br
\end{tabular} 
\end{center}
\end{table}

\section{Efficient computation of loosely coherent sums}
We will now turn to efficient computation of the loosely coherent statistic. Given reduced sensitivity to perturbations in search parameters compared with purely coherent methods and corresponding reduction in the number of templates,  the quadratic cost of computing the sum (\ref{eqn:kernel_statistic}) is not completely unreasonable.

Noticing that the kernel $K_{lm}$ is a positive symmetric matrix, one expects to do better by finding eigenvectors and eigenvalues of $K$ and discarding eigenvectors with small eigenvalues. This will make the computational cost bilinear in $N$ and the number of remaining eigenvectors.

Let us consider, as an example, the case of limited phase evolution $|\phi_k-\phi_{k+1}|<\delta$ with the previously computed kernel
\begin{equation}
\label{eqn:exp_kernel}
K_{lm}=\left(\frac{\sin(\delta)}{\delta}\right)^{|l-m|}=e^{-\alpha |l-m|}
\end{equation}
where we introduced $\alpha=- \log\left(\frac{\sin(\delta)}{\delta}\right)$.

When $\delta=0$ we are dealing with a fully coherent case and the kernel has only one eigenvector with non-zero eigenvalue, while for $\delta=\infty$ we have the semi-coherent case and $K$ is the identity matrix for which we have to use the entire basis. It seems reasonable to expect that for small $\delta$ we will have a few-eigenvector situation, while for large $\delta$ we will have something similar to a semi-coherent sum, where it makes sense not to truncate by eigenvalue but rather 
cut side diagonals of $K$ that are small.

\begin{figure}[htbp]
\begin{center}
  \includegraphics[height=8.0cm]{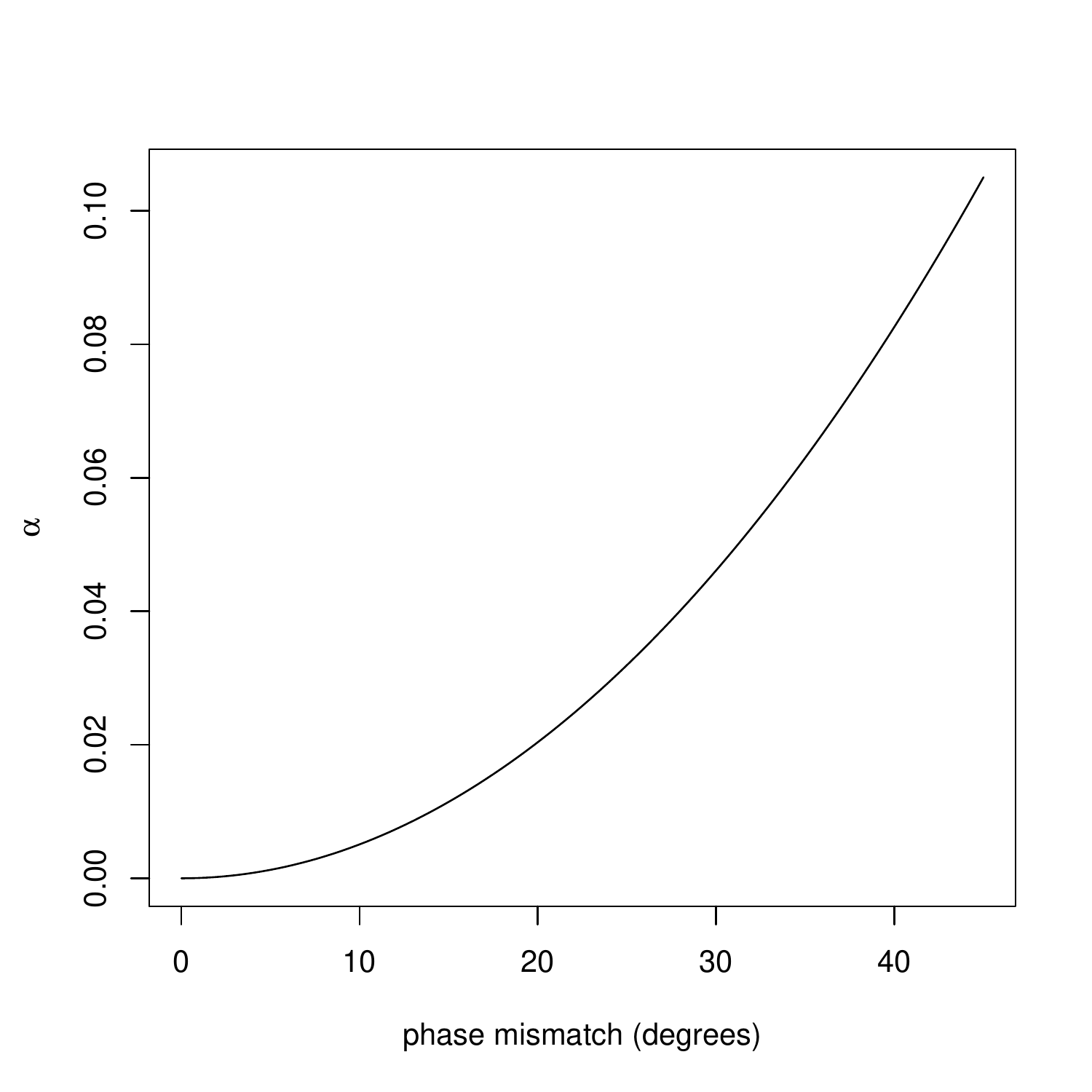}

 \caption[Dependence of $\alpha$ on phase mismatch $\delta$]{Dependence of $\alpha$ on phase mismatch $\delta$}
\label{fig:alpha_delta}
\end{center}
\end{figure}

It turns out that the set of ``small'' $\delta$ values is quite large. To see why this is so, first examine the plot of $\alpha$ versus phase mismatch $\delta$ on figure \ref{fig:alpha_delta}. Even for a phase mismatch as much as $45^\circ$ the value of $\alpha$ is relatively small at $\approx 0.1$.

Secondly, consider the continuous version of our kernel:
\begin{equation}
\left[{\mathbb K} f(v)\right](u)=\int^\infty_{-\infty} e^{-\alpha |u-v|} f(v) dv
\end{equation}
The operator ${\mathbb K}$ is given by a convolution of $f(v)$ with $e^{-\alpha |u|}$. As is well-known, Fourier transform will convert convolution into multiplication. Thus the spectrum of the convolution operator is given as Fourier transform of its kernel. The functions $e^{i\lambda u}$ can be considered as eigenvectors of ${\mathbb K}$ in appropriate functional space (e.g. $C^\infty$):
\begin{equation}
\left[{\mathbb K}e^{i\lambda v}\right](u)=\int^\infty_{-\infty} e^{-\alpha |u-v|} e^{i\lambda v} dv=\frac{2\alpha}{\alpha^2+\lambda^2}e^{i\lambda u}
\end{equation}
The eigenvalues have the familiar Lorentzian form $\frac{2\alpha}{\alpha^2+\lambda^2}$ with quadratic decay. In hindsight, this is not surprising as the condition $|\phi_k-\phi_{k+1}|<\delta$ is  similar to the requirement that the signals we are looking for are band limited.

\begin{table}[htbp]
\caption{Number of eigenvectors required to compute $K$ with $1\%$ accuracy.}
\begin{center}
\lineup
\begin{tabular}{@{}llll}
\br
Phase shift $\delta$ & $N=1000$ & $N=5000$ & $N=10000$ \\
\mr
$1^\circ$ & \0\02  & \0\0\03   & \0\0\04  \\
$5^\circ$ & \0\07  & \0\022  &\0\042  \\
$10^\circ$ & \018  &\0\081  &\0162  \\
$20^\circ$ & \066  & \0324  &\0647  \\
$30^\circ$ &  148  &  \0737  &  1474  \\
$45^\circ$ &  351  &  1751  &  3502  \\
\br
\end{tabular} 
\label{tab:eigencount1}
\end{center}
\end{table}

\begin{table}[htbp]
\caption{Number of eigenvectors required to compute $K$ with $5\%$ accuracy.}
\begin{center}
\lineup
\begin{tabular}{@{}llll}
\br
Phase shift $\delta$ & $N=1000$ & $N=5000$ & $N=10000$ \\
\mr
$1^\circ$ & \0\0$1$ &  \0\0$2$  & \0\0\0$2$ \\
$5^\circ$ & \0\0$3$ &  \0$10$ & \0\0$18$ \\
$10^\circ$ & \0\0$8$ & \0$36$ & \0\0$71$ \\
$20^\circ$ & \0$29$ & $142$ & \0$283$ \\
$30^\circ$ & \0$65$ & $321$ & \0$641$ \\
$45^\circ$ & $147$ & $736$ & $1471$ \\
\br
\end{tabular} 
\label{tab:eigencount5}
\end{center}
\end{table}

Tables \ref{tab:eigencount1} and \ref{tab:eigencount5} show the number of eigenvectors needed to approximate $K$ given by formula \ref{eqn:exp_kernel} for various numbers of equally spaced SFTs and some typical values of $\delta$. The approximation is done using operator norm which is equivalent to counting the number of eigenvalues that are at least 1\% for table \ref{tab:eigencount1} (5\% for table \ref{tab:eigencount5}) of the largest eigenvalue of $K$. While the fraction of eigenvectors does rise linearly with $N$ and thus the computational requirements are still quadratic, said fraction is a rather small number for $\delta \leq 10^\circ$ and in practical implementations (especially on processors with vector arithmetic) the scaling will be close to linear.

For larger values of $\delta$ one might wish to go with a different algorithm. In particular, it makes sense to consider decompositions using non-orthogonal vectors, the simplest of which is obtained by truncation of side diagonals. 

\begin{figure}[htbp]
\begin{center}
  \includegraphics[height=8.0cm]{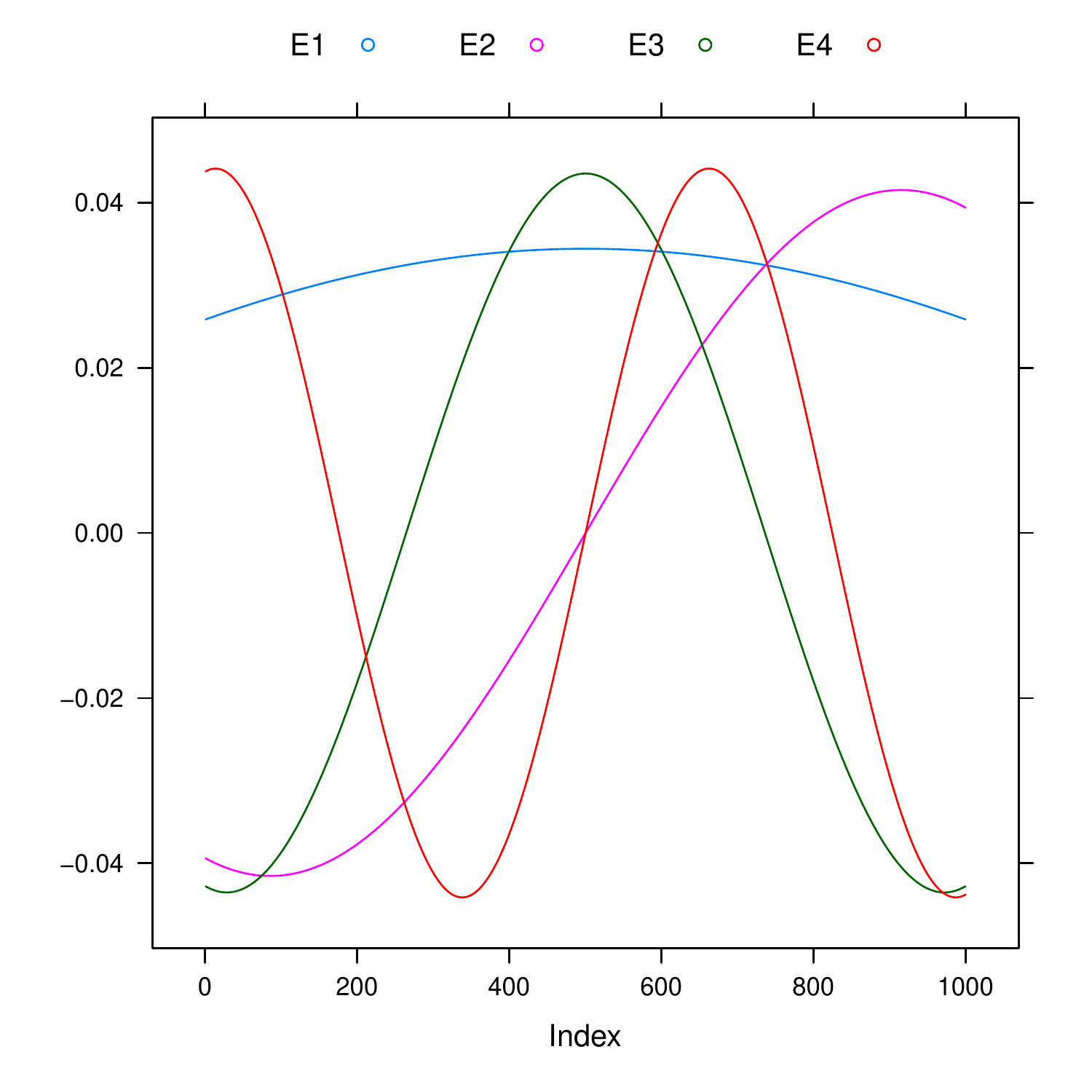}

 \caption[Eigenvectors corresponding to first four largest eigenvalues for $N=1000$ and $\delta=5^\circ$]{Eigenvectors of the kernel \ref{eqn:exp_kernel} corresponding to first four largest eigenvalues for $N=1000$ and $\delta=5^\circ$}
\label{fig:eigenvectors1000}
\end{center}
\end{figure}
 
As we mentioned before, in the continuous case the eigenvectors are simple sine waves $e^{i\omega u}$. The discrete case is nearly sinusoidal. The eigenvectors of the kernel \ref{eqn:exp_kernel} corresponding to first four largest eigenvalues for $N=1000$ and $\delta=5^\circ$ are shown on the figure \ref{fig:eigenvectors1000}. The eigenvector corresponding to the largest eigenvalue is not constant and can be regarded as a window one applies to the data in order to make the usual coherent sum respond to signals from $S$.
The eigenvector decomposition was done numerically using R \cite{R}.

This idea can be exploited to speed up eigenvector decomposition, by analytically transitioning into the basis of pure sine waves and then discarding entries of $K$ from higher order modes. The remaining matrix of smaller dimension can then be diagonalized with conventional numerical techniques.

It is interesting to consider the case of very short Fourier transforms of a few seconds in length and correspondingly small $\Delta t$. The phase shifts from most sources (except for frequency mismatch) will be small as well, and computation of $K$ can be performed by taking a Fourier transform of the input data and then summing up power in low frequency harmonics weighted by eigenvalues of $K$. 

This has close relation to the resampling technique \cite{resampling}.

The resampling implementation of $\mathcal F$-statistic  operates by heterodyning 30 minute SFTs to a desired frequency, inverting the Fourier transform to obtain a time series which is stitched together and then band-limited and downsampled. 
The resulting time series is converted into detector frame which allows efficient computation of $\mathcal F$-statistic using Fourier transform.

Another way to obtain the same time series is to start with shorter SFTs which frequency bins are large enough to accommodate Doppler shift. A time series of frequency bins of these short SFTs is then just another way of heterodyning our input data with the advantage of bypassing the need for inverse Fourier transform. If the frequency band that is being searched is significantly smaller than the size of initial frequency bins the time series can be band-limited and downsampled just as done in \cite{resampling}.

The conversion of heterodyned time-series into detector frame consists of two parts: removal of the phase shift from signal evolution due to intrinsic effects or Earth motion, which is also done by loosely coherent method, and interpolation in order to obtain evenly spaced time series suitable for fast Fourier transform algorithm. 

The computation of $\mathcal F$-statistic involves summing three terms quadratic in the elements of our time series with coefficients that depend on time position of the source and the detector but not the amplitude or polarization of the expected signal. This can be viewed as computation of a specific kernel $K$ which rank is at most $2$. If we take the interpolation algorithm into account the rank will increase but will still be much smaller than kernel dimension.

The same approach can be used to compute loosely coherent statistic where we might need to use additional terms to accommodate kernels with larger rank. In return, the statistic can be made more tolerant of mismatch in source parameters, such as sky location.

\section{Sensitivity estimates}
It must be said that the sensitivity of a given method is best judged from a search made on real data, as computational efficiency and practicalities of detector artifacts in the input data have often a much stronger impact than an extra few percent gained by fine-tuning the algorithm with analytical considerations that assume Gaussian noise.

Nevertheless, it is useful to have an idea of what to expect in the perfect situation as a starting point for practical applications.
We will concentrate on the case of perfectly coherent signal and how the performance varies between the extremes of coherent and semi-coherent power sums.

The standard methods of filtering theory can be employed to obtain a rough estimate. As we mentioned before, the phase evolution condition $|\phi_k-\phi_{k+1}|< \delta$ is closely related to the condition that our signals are band limited. In this case, the rejection of noise outside the acceptance band results in improvement in the signal-to-noise ratio compared to the usual semi-coherent case which is sensitive to all signals within the frequency bin of the original SFTs. 

The acceptance band is narrowed down by a factor inversely proportional to the number of SFTs it takes for the phase to make a full turn (not to exceed, of course, the total number of SFTs available). Thus, given a fixed number of SFTs, we expect the improvement in the signal-to-noise ratio to scale as $1/\sqrt{\delta}$ tempered by the non-linear effects of our statistic.

\begin{figure}[htbp]
\begin{center}
  \includegraphics[height=8.0cm]{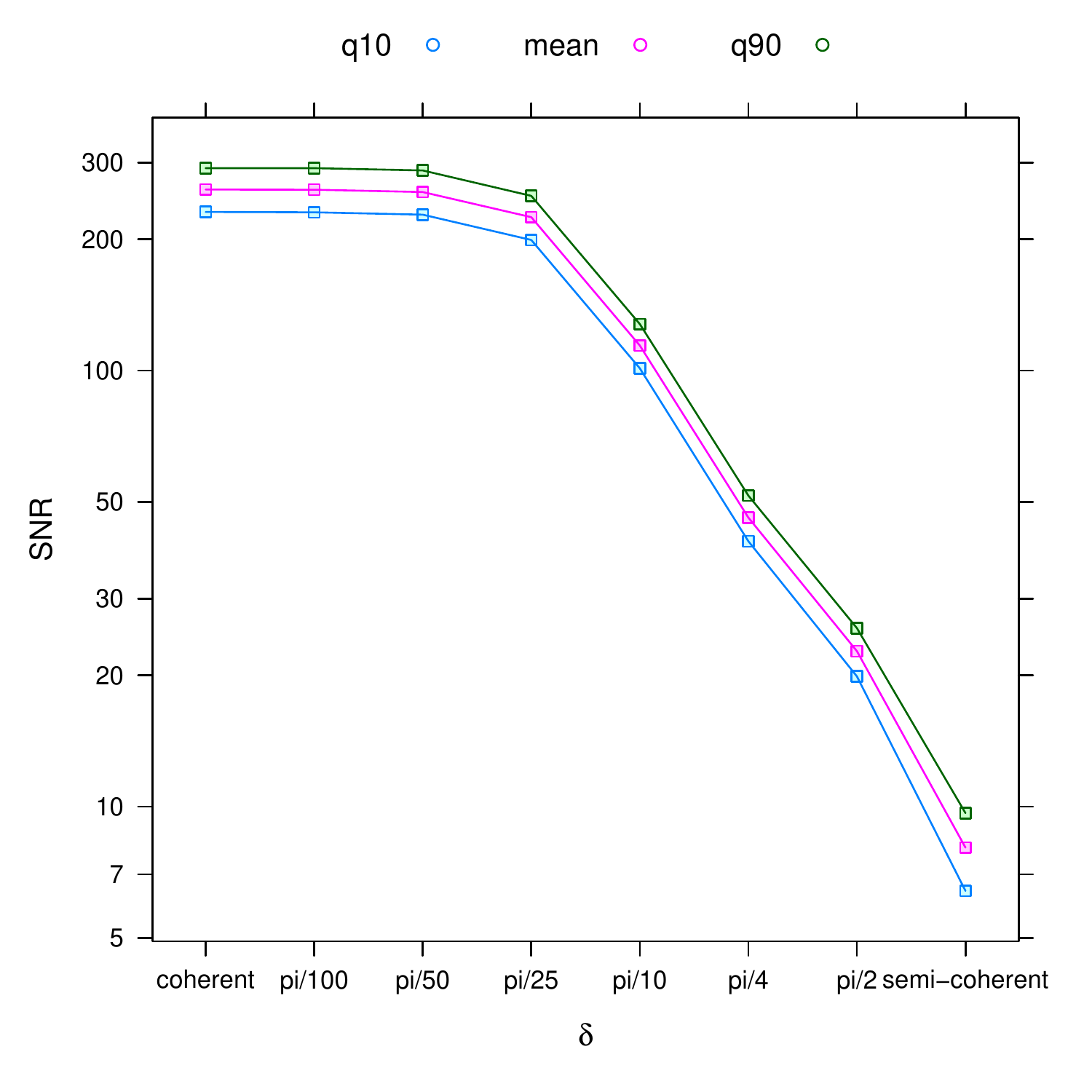}

 \caption[Dependence of signal-to-noise ratio on phase mismatch $\delta$]{Dependence of signal-to-noise ratio on phase mismatch $\delta$. The upper, central and lower curves show 90\% quantile, mean and 10\% quantile of multiple simulation runs. }
\label{fig:loose_SNR}
\end{center}
\end{figure}

This is illustrated on figure \ref{fig:loose_SNR} that shows results of simulation evaluating signal-to-noise ratio gain for limited phase evolution statistic as we decrease $\delta$ for a coherent signal. The simulation was performed using $N=1000$ SFTs which were composed of Gaussian noise $\xi_i$ with standard deviation $1$ and a constant signal with amplitude $h=0.7$ which results in the average signal-to-noise ratio of $\approx 8$ for a semi-coherent search. 

The statistic was computed according to the formula
\begin{equation}
P(h)=\sum_{i,j=1}^{N} K_{ij}(\xi_i+h)(\bar{\xi}_j+\bar{h})
\end{equation}
where the kernel $K_{ij}$ was either an identity matrix for semi-coherent case, a matrix with $1$ in all cells for the coherent case or given by the formula \ref{eqn:exp_kernel} for the loosely coherent case.

The signal-to-noise ratio in this simulation was defined as the value of the statistic minus the average value obtained on noise alone and divided by the standard deviation of values produced by pure noise:
\begin{equation}
\SNR=\frac{\mathrm{mean} (P(h))- \mathrm{mean}( P(0))}{{\mathrm{sd}} (P(0))}
\end{equation}
Here mean and standard deviation were taken over $1000$ independent realizations of noise.

All of the statistic values are described by a weighted $\chi$-squared distribution which depends on $\delta$. For large $\delta$, however, it is close to a Gaussian distribution as well due to the central limit theorem. To illustrate the change in the distribution of our statistic we show
$10\%$ and $90\%$ quantiles of the signal-to-noise ratios obtained as well as the mean. The vertical axis is logarithmic, so the spread in signal-to-noise ratios increases as $\delta$ becomes smaller.

The flattening out of the curve for small $\delta$ is due to different scaling regimes near the extremes of coherent and semi-coherent statistics. This can be illustrated by considering a semi-coherent statistic that operates on $N$ SFTs which are coherently combined in stretches of $k$ SFTs each and the results are combined incoherently. Then the scaling law for the signal-to-noise ratio is:
\begin{equation}
\SNR\sim k\sqrt{N/k}
\end{equation}
Now suppose that $k=\alpha N$ is a certain fraction of $N$. Then the scaling is
\begin{equation}
\SNR\sim N\sqrt{\alpha}
\end{equation}
As our statistic is power based the sensivity will scale as $1/(\sqrt[4]{\alpha}\sqrt{N})$.
The fourth root in $\alpha$ has a really slow growth. For example, for $\alpha=0.1$ it is only $0.56$ - so for less than a factor of $2$ loss in sensitivity the coherence length can be dropped by a factor of $10$.

\section{Prototype implementation}
An initial implementation of the loosely coherent statistic was done within the framework of the PowerFlux \cite{PowerFluxTechNote} program. This implementation provided practical experience with a loosely coherent search and addressed the problem of following up outliers from the all-sky PowerFlux search over LIGO's fifth science run.

As the underlying code base was not designed with the loosely coherent search in mind, the code has a number of inefficiencies. In particular, the double sum in the statistic $\expect P$ was computed by brute force. Nevertheless, the speed was sufficient to quickly carry out searches in disks of $0.03$ radians radius on the sky over $20002$ SFTs split evenly between H1 and L1 detectors. The powers from individual detectors were combined incoherently to make the comparison to semi-coherent code more fair. The nearby SFTs were separated by $30$~min. In practical data, the SFTs are usually $50\%$ overlapped, but there are can also be gaps in the data. The $30$~min constant was chosen as a reasonable worst case.

While the analysis of actual interferometer data is still underway, we can report on results of simulations using Gaussian data. For these simulations we used a Lanczos kernel with parameter $3$:
\begin{equation*}
K(t_1, t_2)=\left\{
\begin{array}{ll}
\frac{\sin(\delta(t_2-t_1)/30\textrm{~min})\sin(\frac{1}{3}\delta(t_2-t_1)/30\textrm{~min})}{\delta^2((t_2-t_1)/30\textrm{~min})^2} &  \textrm{\quad when~}\frac{\delta\left|t_2-t_1\right|}{30\textrm{~min}}<3\pi\\
0 & \textrm{\quad when~} \frac{\delta\left|t_2-t_1\right|}{30\textrm{~min}}\ge 3\pi\\
\end{array}
\right.
\end{equation*}
This kernel naturally vanishes for widely separated SFTs which makes this a variant of cross-correlation search, albeit with particularly large number of off-diagonal entries, which is further increased by the $50\%$ overlap of nearby SFTs that is usually employed by PowerFlux. We explored values of $\delta$ as small as $\pi/5$ which involves summing up to $59$ diagonals when working with overlapped SFTs. For these values of $\delta$ the required computational time scales as square of observation time (for time bases several months and larger) and
as a cube of covered frequency range.  

\begin{figure}[htbp]
\begin{center}
  \includegraphics[height=8.0cm]{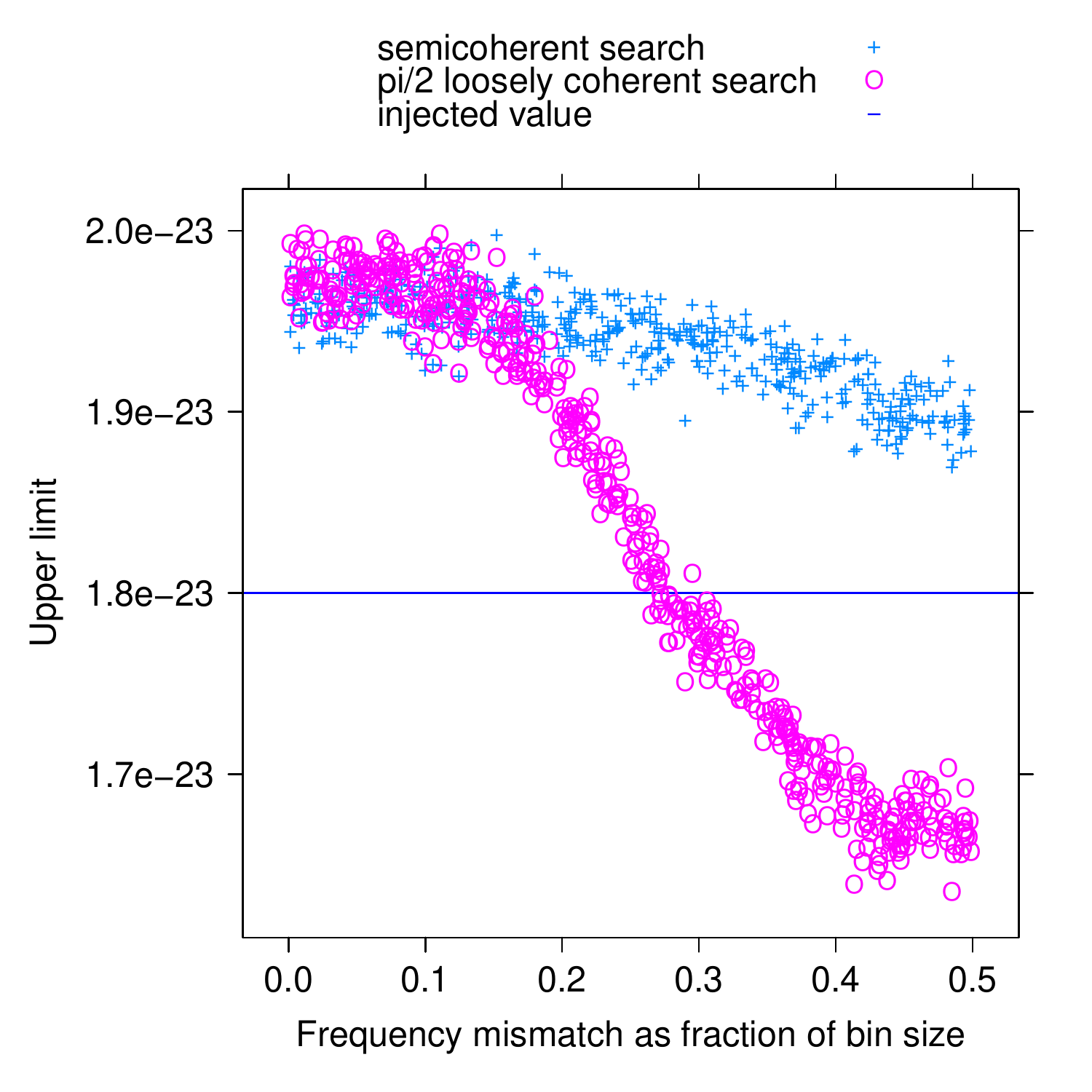}

 \caption[Upper limit versus frequency bin mismatch.]{Upper limit versus frequency bin mismatch. The horizontal line marks the strain of the software injections. The upper curve shows upper limits from a semi-coherent search which are consistently above injected value. The upper limits from loosely coherent search follow semi-coherent search before the limit of phase tolerance is reached and decline sharply afterwards. }
\label{fig:h0_vs_fbin}
\end{center}
\end{figure}

\begin{figure}[htbp]
\begin{center}
  \includegraphics[height=8.0cm]{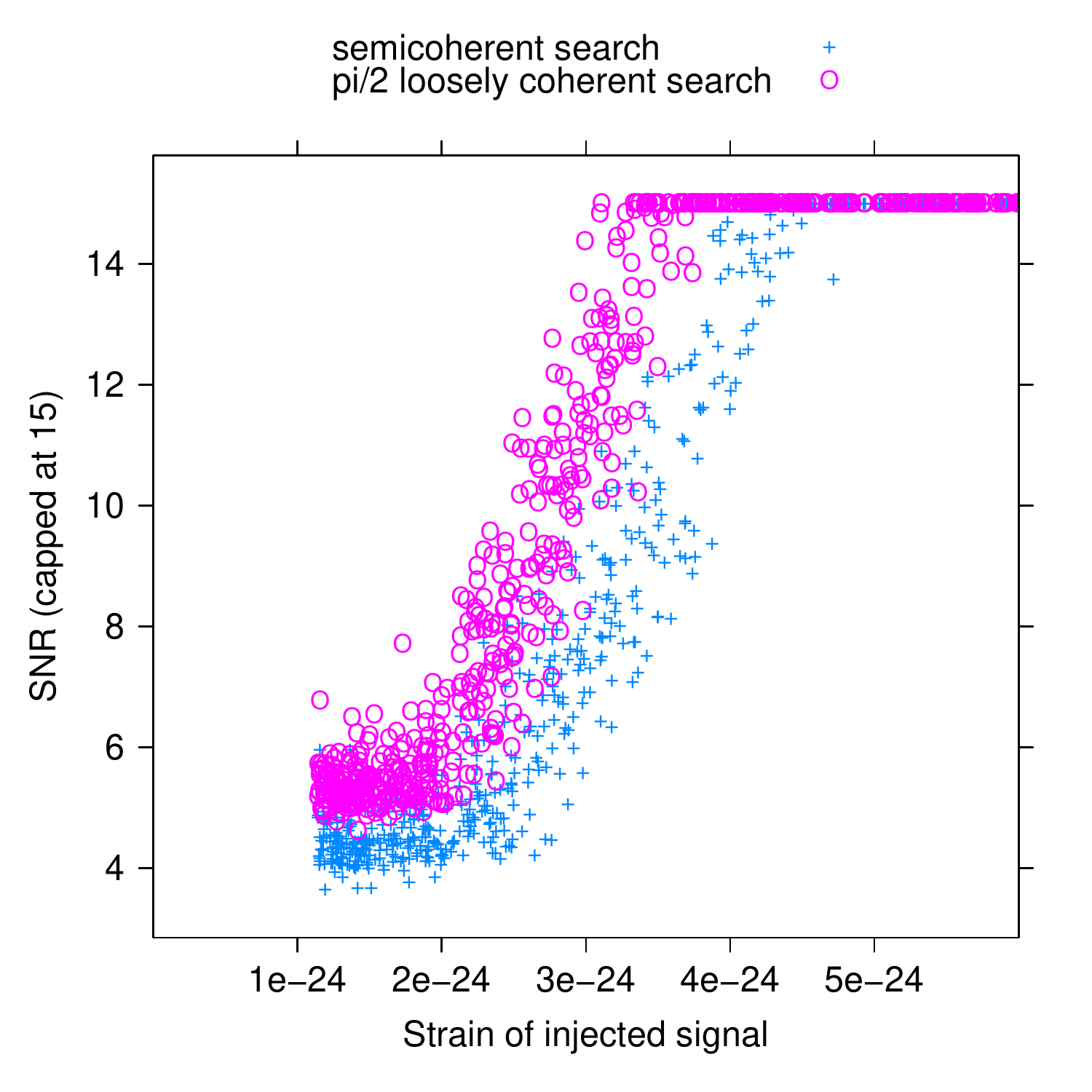}

 \caption[SNR versus injected strain]{Signal to noise ratio versus strain of Monte-Carlo injections. The values were capped at 15 in order to expose the more interesting low SNR region.}
\label{fig:snr_vs_h0}
\end{center}
\end{figure}

Figures \ref{fig:h0_vs_fbin} and \ref{fig:snr_vs_h0} show results of Monte-Carlo injection run assuming a static source location (right ascension 2.0, declination 1.0, spindown 0) and a linearly polarized signal. This choice was made to increase readability of the plots as all-sky injections with arbitrary polarizations inject different amount of power in the interferometer making the curves wider. The injections were made into Gaussian data that was filtered to simulate Hann windowed short Fourier transforms (SFTs). The assumed frequency range varied from $400$ to $410$~Hz and SFT frequency bin size was $1/1800$~Hz.

The 95\% confidence level upper limits are produced by PowerFlux code for a set of 501 frequency bins given a particular direction on the sky and a spindown value. The results are then maximized over a set of polarizations and small area on the sky around the injection point. This follows the analysis method used in \cite{S4IncoherentPaper} and \cite{EarlyS5Paper}.

Both semi-coherent (power only) and loosely coherent algorithms proceed by sampling discrete range of frequencies with configurable spacing in fractions of SFT bin size. Figure \ref{fig:h0_vs_fbin} compares how the mismatch between the actual injected frequency and the sampled frequency affects upper limits produced by semi-coherent and loosely coherent codes. The frequency spacing was set at $1$~SFT~bin and the injected strain value was fixed to $\sci{1.8}{-23}$. We see that a loosely coherent search with $\delta=\pi/2$ has an initial flat response for small mismatch in frequency which is followed by rapid decay to values below injected strain. In contrast, the semi-coherent search shows only minor reduction in the upper limit which is fully compensated by built-in correction factor.

Figure \ref{fig:snr_vs_h0} compares the signal to noise ratios (SNRs) of semi-coherent and loose-coherent methods. The frequency spacing of the loosely coherent search was reduced to $1/8$th of the SFT bin which insures correct reconstruction of the upper limit
for the entire range of weak and strong signals. Because of the larger number of templates, the SNR achieved on pure noise is higher for the loosely coherent search than that of the semi-coherent search.
For signals above noise the loosely coherent search produces signal-to-noise ratios on average $50$\% larger than semi-coherent one.

\section{Summary}
We have discussed the problem of detecting a family of signals $S$ from the point of view of computational efficiency and presented a method of creating a statistic that is sensitive to the entire family $S$ or its subset. Two simple examples were considered which showed close ties to well-known methods of matched filtering, cross-correlation and semi-coherent sums.

There are several directions of further study:
\begin{itemize}
 \item The prototype large $\delta$ implementation shows feasibility of the overall method, but does not provide information on the overall computational efficiency. We plan to develop a dedicated small $\delta$ code to be used in targeted searches that cover small sky area (such as galactic center or globular clusters). This should provide experience with scalability properties of the loosely coherent method.
 \item The average of $P$ was used to make the maximization computationally tractable. In fact, for small $N$ the maximization can be carried out directly. It is worthwhile to investigate the possibility of combining the two techniques.
 \item For the case of the set $S$ given by conditions $|\phi_k-\phi_{k+1}|<\delta$ and assuming small $\delta$ the maximization over $P$ can be carried out assuming $\phi_{k+1}=\phi_k\pm \delta$. This converts the problem into the discrete domain and makes it amenable to binary optimization methods which have seen much progress in recent years. A particularly interesting observation is that for a noise dominated signal the function to be optimized has random coefficients, so an optimization method that works only on a certain proportion of objective functions can yield useful results.
\end{itemize}

\section*{Acknowledgments}
This work has been done while being a member of LIGO laboratory, supported by funding from United States National Science Foundation. The simulations were completed on the wonderful ATLAS cluster at Albert Einstein Institute, with special thanks due to Bruce Allen, Carsten Aulbert, Henning Fehrmann and Miroslav Shaltev. The author has greatly benefited from discussions with his colleagues, in particular  Joe Betzweiser, Chris Messenger and Keith Riles. We are greatly thankful to the referee for many useful comments and suggestions.
This document has the LIGO document number P1000015.


\end{document}